\documentclass[%
 reprint,
amsmath,amssymb,
aps,
]{revtex4-2}

\usepackage{graphicx}
\usepackage{dcolumn}
\usepackage{bm}
\usepackage{orcidlink} 
\usepackage{float}

\begin{document}

\preprint{APS/123-QED}

\title{Sensitivity to CP Discovery in the Presence of Lorentz Invariance Violating Potential at T2HK/T2HKK}
\author{Supriya Pan$^{1,2}$\orcidlink{0000-0003-3556-8619}}
\email{supriyapan@prl.res.in}
\author{Kaustav Chakraborty$^1$\orcidlink{0000-0002-1847-8403}}
\email{kaustav.chk@gmail.com}
\author{Srubabati Goswami$^{1,3}$\orcidlink{0000-0002-5614-4092}}%
\email{sruba@prl.res.in}

\affiliation{$^1$ Physical Research Laboratory, Ahmedabad, Gujarat, 380009, India}
\affiliation{$^2$ Indian Institute of Technology, Gandhinagar, Gujarat, 382355, India}
\affiliation{$^3$ Northwestern University, Department of Physics and Astronomy, Evanston, IL 60208, USA}
%

%



\begin{abstract}
Investigation of conservation/violation of CP symmetry in the leptonic sector is very essential in understanding the evolution of the universe. Lorentz invariance and CPT are fundamental symmetries of nature. The violation of Lorentz invariance can also lead to CPT violations. The standard three flavour neutrino oscillation framework presents a scenario to observe the signature of Lorentz invariance and CP violations. This work focuses on the effect of Lorentz invariance violating (LIV) parameters on the sensitivity to CP violation. We investigate the sensitivity in two proposed configurations of the upcoming T2HK experiment: (i) one detector each placed at 295 km and 1100 km, and (ii) two identical detectors at 295 km. This study probes the effect of CPT violating parameters $a_{e\mu},a_{e\tau},a_{\mu\tau}$.
\end{abstract}

\maketitle


\section{\label{sec:intro}Introduction}

The standard model (SM) of particle physics describes the elementary particles and their interactions except gravity. Its effectiveness has been tested thoroughly over the years in a successful manner. However, there have been a few shortcomings that are essential to address. Neutrino oscillation is one such phenomenon in which there is inter-conversion among three flavors $\nu_e, \nu_{\mu}, \nu_{\tau}$. This can occur only if the neutrinos are massive. However, in the SM, neutrinos are massless particles. Incorporation of neutrino mass requires some new physics beyond the standard model (BSM). The scale of new physics may be at a higher energy scale, and the standard model can be regarded as an effective low-energetic version of the high energy theory.

The parameters of the three neutrino oscillation framework are: mixing angles $\theta_{12},\theta_{13},\theta_{23}$ and Dirac CP phase $\delta_{13}$ along with the two mass squared differences $\Delta_{31}=m_3^2-m_1^2,\Delta_{21}=m_2^2-m_1^2$ corresponding to mass eigenstates $m_1,m_2,m_3$. Currently, the precise measurement of the octant of $\theta_{23}$, the CP phase $\delta_{13}$, and the sign of atmospheric mass squared difference $\Delta_{31}$ is still awaited and probed by current oscillation experiments. Several future experiments with enhanced capabilities are in the pipeline to measure these parameters with increased precision. They also open up the possibility of testing different BSM scenarios. Probing new physics beyond the standard model, like sterile neutrinos, non-standard interactions, neutrino decay, long range forces, CPT and Lorentz invariance violations, etc., in these experiments, is currently a very active area of research in neutrino physics.

In this work, the new physics that we focus on is the CPT non-conserving Lorentz invariance violation. The local relativistic quantum field theories form the basic structure of nature. The primary assumptions of this theory consist of Lorentz invariance, locality of the interactions, along with the hermicity of the Hamiltonian. The conservation of CPT symmetry is embedded into these assumptions. A spontaneous violation of CPT will always be associated with Lorentz invariance violation but not vice versa. Since both CPT and Lorentz invariance are fundamental symmetries of nature, any violation of them will provide us with an indication of new physics governing the laws of nature. 

Lorentz invariance protects isotropy and homogeneity of the local relativistic QFT in space-time. In the minimal $SU(3)\times SU(2)\times U(1)$ SM, this symmetry is conserved. However, there are higher dimensional theories (related to the Plank scale $\sim 10^{19}$ GeV) where Lorentz invariance violation is generated spontaneously\cite{Kostelecky:1988zi, Kostelecky:1989jp, Kostelecky:1989jw, Kostelecky:1990pe, Colladay:1996iz}. String theories can give rise to the spontaneous breaking of Lorentz symmetry \cite{Kostelecky:1988zi, Kostelecky:1989jp, Kostelecky:1989jw, Kostelecky:1990pe}. At the level of the standard model, Lorentz invariance violation can be manifested as a new physics effect suppressed by the Planck scale. There is also a proposed effective field theoretical description, as given in ref. \cite{Colladay:1996iz} that will be consistent with the known physics phenomena of the SM at low energy. The violation of Lorentz invariance and CPT have been tested using Kaons \cite{Kostelecky:1991ak, Kostelecky:1994rn}, neutral $\rm{B_d}$ or $\rm{B_s}$ mesons \cite{Kostelecky:1994rn, Colladay:1994cj}, and neutral D mesons \cite{Kostelecky:1994rn, Colladay:1995qb}. It has also been realized that neutrino oscillation experiments can provide a testing ground for probing signatures of LIV, and there have been several studies on the implications of LIV for various experiments \cite{DoubleChooz:2012eiq, MiniBooNE:2011pix, MINOS:2008fnv, MINOS:2010kat, IceCube:2010fyu, LSND:2005oop, SNO:2018mge, Huang:2019etr}.

The presence of LIV will also modify the standard neutrino oscillation probabilities, and this can impact the determination of the standard $3\nu$ parameters in future neutrino oscillation experiments. This aspect and the constraints on the LIV parameters have been investigated in several recent works in the context of accelerator and atmospheric neutrino experiments. In ref.~\cite{Raikwal:2023lzk}, a study has been performed to give bounds on LIV parameters using INO-ICAL, T2HK, and DUNE. There has been a study on the effect of LIV parameters in NO$\nu$A and T2K in ref.~\cite{Rahaman:2021leu}. Efforts have been made to separately understand the effects of LIV interactions and non-standard interactions (NSI) at long baseline experiments in \cite{Majhi:2022fed, Sahoo:2023gcf}. Other recent studies related to  CPT violation and LIV interactions in neutrinos can be found in \cite{KATRIN:2022qou, Fiza:2022xfw, Li:2022sgs, Skrzypek:2023blj, Li:2023wlo, KumarAgarwalla:2019gdj, Majhi:2019tfi, EXO-200:2016hbz, SNO:2018mge}.

In our study, we focus on the effects of CPT violating LIV parameters on the determination of the CP phase in the upcoming T2HK\cite{Hyper-Kamiokande:2016dsw}/T2HKK\cite{Hyper-Kamiokande:2016srs} detector. T2HK plans to have two detectors at baseline 295 km (first oscillation maxima) and T2HKK is proposed to have one detector each at 295 km and 1100 km(second oscillation maxima). We compare and contrast the CP sensitivities in the presence of LIV at both these baselines and explore the synergistic effect between the first and second oscillation maxima.
 
The structure of the paper is as follows. At first, the formalism of LIV in the neutrino sector is described in \autoref{sec:LIV-th}. It's followed by a discussion on the dependence of the neutrino oscillation probabilities on LIV parameters in \autoref{sec:LIV-prob}. The numerical analysis for CP discovery in the presence of non-diagonal CPT violating LIV parameters is presented in \autoref{sec:chi}. In \autoref{sec:precision-LIV}, the precision of the LIV parameters is discussed. Finally, we conclude in \autoref{sec:result}.

\section{\label{sec:LIV-th}Theory of Lorentz invariance violation}

Lorentz invariance violation(LIV) can be comprehended through a standard model extension (SME) framework in the context of a low energy effective theory\cite{PhysRevD.55.6760}. The neutrino behaviour is contained in the following Lagrangian,
\begin{equation}\label{eq:liv_lag}
	\mathcal{L} =\frac{1}{2}\iota \bar{L}_a \gamma^\mu \overleftrightarrow{D_\mu} L_a - (a_L)_{\mu a b}\bar{L}_a \gamma^\mu L_b + \frac{1}{2}(c_L)_{\mu\nu a b}\bar{L}_a\gamma^\mu \overleftrightarrow{D^\nu} L_b
\end{equation}
where the first term is the usual Standard-Model kinetic term for the left-handed doublets $L_a$ with index $a$ ranging over the three generations $e,\mu,\tau$. The coefficients for Lorentz violation are $(a_L)_{\mu a b}$, which has mass dimension one and controls the CPT violation, and $(c_L)_{\mu \nu a b}$ which is dimensionless and is CPT conserving. The Lorentz-violating terms in eq.~\eqref{eq:liv_lag} modify both interactions and propagation of neutrinos. Any interaction effects are expected to be tiny and well beyond the existing sensitivities. In contrast, propagation effects can be substantial if the neutrinos travel large distances. The time evolution of neutrino states is controlled by the effective Hamiltonian extracted from eq.~\eqref{eq:liv_lag} as,
\begin{equation}
	(\mathcal{H}_{eff})_{ab}= E \delta_{ab} + \frac{m^2_{ab}}{2E} + \frac{1}{E}(a_L^\mu p_\mu -c_L^{\mu\nu}p_\mu p_\nu)_{ab}
\end{equation}
The LIV-induced parameter $ a_L^\mu $ (CPT-violating) will change the sign in case of anti-neutrinos while $ c_L^{\mu\nu} $ will remain unchanged. Here we will focus on only the isotropic component of these parameters in the Sun-centered celestial-equatorial frame and fix $ \mu,\nu$ to be zero and redefine $ (a_L)^0_{ab} \equiv a_{ab},(c_L)^{00}_{ab}\equiv c_{ab} $. The Hamiltonian due to LIV is given by,
\begin{equation}
	H_{LIV}=\begin{bmatrix}
		a_{ee}&a_{e\mu}&a_{e\tau}\\
		a^\star_{e\mu}&a_{\mu\mu}&a_{\mu\tau}\\
		a^\star_{e\tau}&a^\star_{\mu\tau}&a_{\tau\tau}\end{bmatrix}-
	\frac{4}{3}E \begin{bmatrix}
		c_{ee}&c_{e\mu}&c_{e\tau}\\
		c^\star_{e\mu}&c_{\mu\mu}&c_{\mu\tau}\\
		c^\star_{e\tau}&c^\star_{\mu\tau}&c_{\tau\tau}
	\end{bmatrix}
\end{equation}
The total Hamiltonian for neutrino propagation, including the standard MSW matter effect, is given by
\begin{equation}
	H_{tot}=\frac{1}{2E}\begin{pmatrix}
		m_1^2&0&0\\0&m_2^2&0\\0&0&m_3^2
	\end{pmatrix}+\begin{pmatrix}
		\sqrt{2}G_F N_e&0&0\\0&0&0\\0&0&0		
	\end{pmatrix}+H_{LIV}
\end{equation}
Here we only consider CPT-violating LIV parameters $ a_{\alpha\beta} $. The non-diagonal parameters are complex and given by $a_{\alpha\beta}=|a_{\alpha\beta}|e^{\iota\phi_{\alpha\beta}} $ where as diagonal parameters $ a_{\alpha\alpha} $ are real.
There is an established correlation between CPT-violating LIV parameters and matter NSI parameters given by
\begin{equation}	\epsilon_{\alpha\beta}^m\equiv\frac{a_{\alpha\beta}}{\sqrt{2}G_F N_e}
\end{equation}
Irrespective of this mapping, their origins are very different as well as their implications, e.g., matter NSI have similar effects as the MSW matter potential in neutrino propagation, whereas CPT-violating LIV has an intrinsic effect on neutrino propagation even in the vacuum \cite{Sahoo:2022nbu}. The current constraints of the CPT violating LIV parameters are given below in table \ref{table:nsi-bound}\cite{Super-Kamiokande:2014exs, IceCube:2017qyp, Abe_2015}.
\begin{table}[h]
	\centering
	\begin{tabular}{|c|c|c|}
		\hline
		Parameter & SK Bound & IceCube Bound\\\hline
		$a_{e\mu}$ & $1.8\times10^{-23}$ GeV & N.A.\\
		$a_{e\tau}$ & $4.1\times10^{-23}$ GeV& N.A. \\
		$a_{\mu\tau}$ & $0.65\times10^{-23}$ GeV & $0.29\times10^{-23}$ GeV\\
		\hline
	\end{tabular}
	\caption{The table depicts $95\%$ C.L. bounds of CPT violating non-diagonal LIV parameters from SK and IceCube experiments}
	\label{table:nsi-bound}
\end{table}
\section{\label{sec:LIV-prob} Probabilities in presence of LIV parameters}
In our study, we probe the effects of the CPT violating LIV parameters $a_{e\mu}, a_{e\tau}, a_{\mu\tau}$ on the discovery of the CP phase. For the scope of this work, we use the proposed long baseline experiment configuration of Tokai to hyper-Kamiokande (T2HK) and Tokai to hyper-Kamiokande and Korea (T2HKK). The setup of T2HKK provides an advantage of two detectors; one at Hyper Kamiokande site, 295 Km away from the source, and another at 1100 km away from the source. At the leading order of $\alpha=\Delta_{21}/\Delta_{31}$, the appearance probability $P_{\mu e}$ depends only on parameters $a_{e\mu}, a_{e\tau},\phi_{e\mu}, \phi_{e\tau}$ whereas, the disappearance probability depends on $a_{\mu\tau}, \phi_{\mu\tau}$. The probabilities are calculated in ref. \cite{Masud:2018pig, Kopp:2007ne} as follows,
  \begin{eqnarray}
	P_{\mu e}=P_{\mu e}^{3\nu} + P_{\mu e}^{a_{e\mu}} + P_{\mu e}^{a_{e\tau}} \label{eq:pme-tot}\\
	P_{\mu \mu}=P_{\mu\mu}^{3\nu} + P_{\mu\mu}^{a_{\mu\tau}} \label{eq:pmm-tot},
\end{eqnarray}
where $P_{\mu e}^{3\nu}, P_{\mu \mu}^{3\nu}$ are the three flavour oscillation probabilities in the matter and the LIV induced part of the probabilities are given as,
\begin{widetext}
    \begin{gather}
         P_{\mu e}^{3\nu}=4 s_{13}^2 s_{23}^2 \frac{\sin^2 [(\hat{A}-1)\Delta]}{(\hat{A}-1)^2}+ 2 \alpha s_{13} \sin{2\theta_{12}}\sin{2\theta_{23}} \frac{\sin[\hat{A} \Delta]}{\hat{A}} \frac{\sin [(\hat{A}-1)\Delta]}{\hat{A}-1} \cos(\Delta+\delta_{13})\\
        P_{\mu\mu}^{3\nu}= 1- \sin^2{2\theta_{23}} \sin^2{\Delta}
    \end{gather}
  \begin{eqnarray}
	P_{\mu e}^{a_{e\mu}} \simeq \frac{4|a_{e\mu}| \hat{A}\Delta s_{13} \sin 2\theta_{23} \sin \Delta}{\sqrt{2} G_F N_e} [Z_{e\mu} \sin(\delta_{13}+\phi_{e\mu}) + W_{e \mu}\cos(\delta_{13}+\phi_{e\mu})] \label{eq:pme-aem}\\
	P_{\mu e}^{a_{e\tau}} \simeq \frac{4|a_{e\tau}| \hat{A}\Delta s_{13} \sin 2\theta_{23} \sin \Delta}{\sqrt{2} G_F N_e} [Z_{e\tau} \sin(\delta_{13}+\phi_{e\tau}) + W_{e \tau}\cos(\delta_{13}+\phi_{e\tau})] \label{eq:pme-aet}\\
	P_{\mu\mu}^{a_{\mu\tau}}=\frac{4|a_{\mu\tau}|\hat{A}\Delta \sin 2\theta_{23}\sin\Delta}{\sqrt{2}G_F N_e}[Z_{\mu\tau}\cos\phi_{\mu\tau}+ W_{\mu\tau}\cos\phi_{\mu\tau}]\label{eq:pmm-amt}
\end{eqnarray}
where $ \Delta=\frac{\Delta_{31}L}{4E} $, $\alpha=\Delta_{21}/\Delta_{31}$, $\hat{A}= \frac{2\sqrt{2}G_F N_e E}{\Delta_{31}} $, $A=2\sqrt{2}G_F N_e E$, $ s_{ij}=\sin\theta_{ij} $, $ c_{ij}=\cos\theta_{ij} $,
\begin{gather}
	Z_{e\mu}=-\cos \theta_{23} \sin\Delta, \hspace{3mm} Z_{e\tau}=\sin \theta_{23} \sin\Delta, \hspace{3mm} Z_{\mu\tau}=-\sin^2 2\theta_{23} \cos\Delta \label{eq:zem-zet-zmt}\\
	W_{e\mu}=c_{23} (\frac{s^2_{23} \sin\Delta}{\Delta. c^2_{23}} + \cos \Delta), W_{e\tau}=s_{23}(\frac{\sin\Delta}{\Delta} - \cos \Delta), W_{\mu\tau}=\frac{-\cos^2 2\theta_{23}\sin \Delta}{\Delta} \label{eq:wem-wet-wmt}
\end{gather}
\end{widetext}
\subsection{\label{subsec:var-Pme}Variation in $ P_{\mu e} $ with phases at fixed $ a_{e\mu}, a_{e\tau}, a_{\mu\tau} $}
\begin{figure}[!h]
	\centering
	\includegraphics[width=0.9\linewidth]{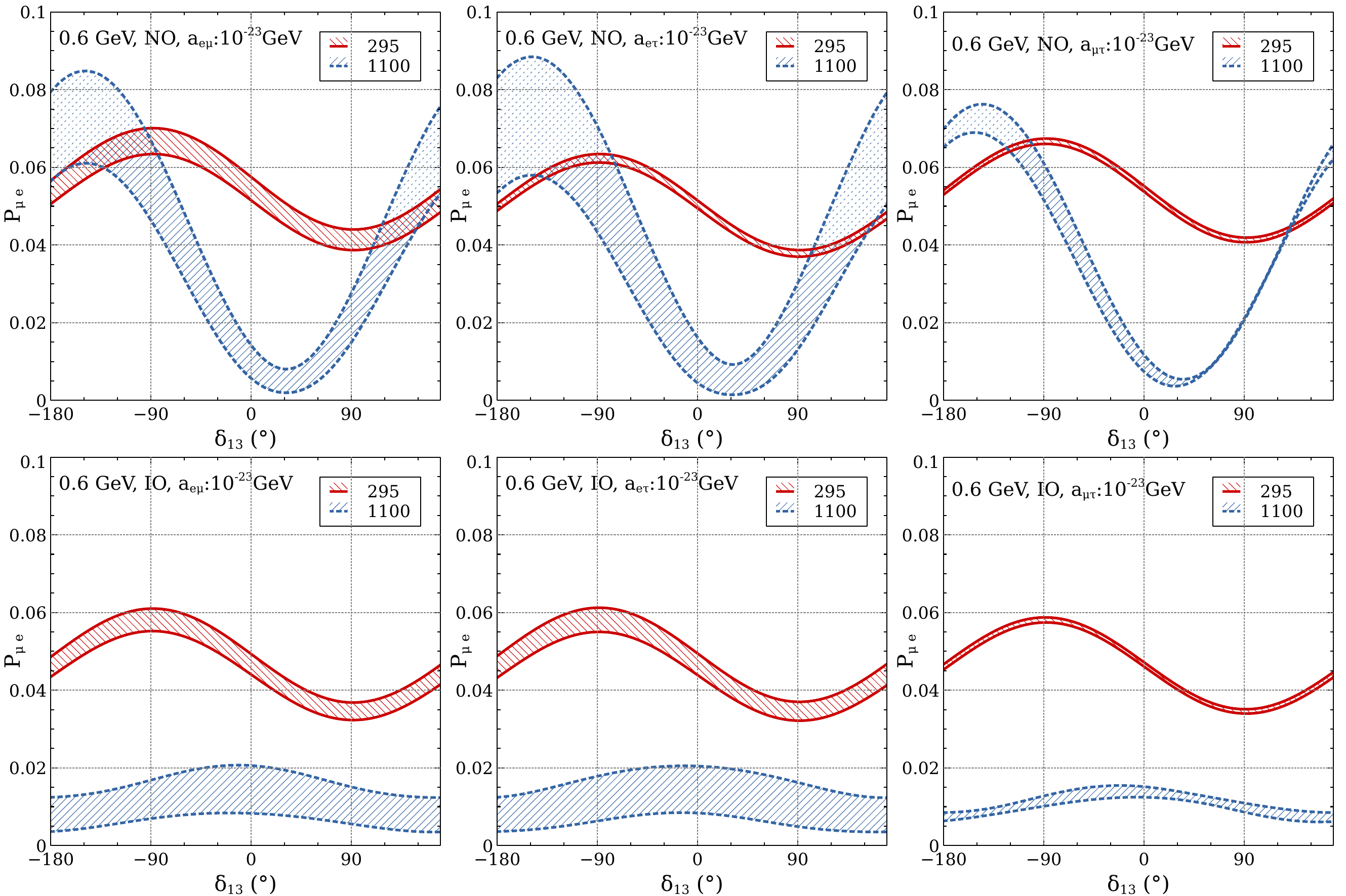}
	\caption{$ P_{\mu e} $ as a function of $\delta_{13}$ for  $\theta_{23}=49^\circ$, $ a_{e \mu}=10^{-23} $ GeV (left),  $a_{e \tau}=10^{-23} $ GeV (middle), and $ a_{\mu \tau}=10^{-23} $ GeV (right) due to variation of respective phases $\phi_{e\mu},\phi_{e\tau},\phi_{\mu \tau}$ for NO(top), IO(bottom) at 0.6 GeV in 295 km (red), 1100 km(blue)}
	\label{fig:pme-dcp} 
\end{figure}

In the presence of the LIV parameters, the appearance channel probability depends on the parameters  $ a_{e\mu}, a_{e\tau}$, and $a_{\mu\tau} $. It also depends on the LIV phases $\phi_{e\mu}, \phi_{e\tau}$ in conjunction with $\delta_{13}$. The modifications in $P_{\mu e}$ due to LIV parameters are probed in this section at 1100 and 295 km baselines. In the plots that follow, the values of the oscillation parameters are chosen as\cite{Esteban_2020},
$\theta_{12}=33.44^\circ, \theta_{13}=8.57^\circ, \theta_{23}=49^\circ,\Delta_{21}=7.42\times10^{-5}$ eV$^2 $, and $|\Delta_{31}|=2.515\times10^{-3}$ eV$^2$. $P_{\mu e}$ is plotted as a function of $\delta_{13}$ at 0.6 GeV in fig.~\ref{fig:pme-dcp} for normal (top panel) and inverted (bottom panel) mass orderings in case of 295 km (red), and 1100 km (blue) baseline while the values of the non-diagonal LIV parameters are kept fixed at $10^{-23}$ GeV. The bands refer to the variation of LIV phases. The significant points to be noted are as follows,

\begin{itemize}
	\item It can be observed from both top and bottom panels that the effect of $\phi_{e\mu}, \phi_{e \tau}$ is larger than $\phi_{\mu\tau}$ as the width of the red and blue bands are narrower in the right panels than the left and middle ones. This can be understood from the eq. \eqref{eq:pme-aem}, \eqref{eq:pme-aet} as $P_{\mu e}$ has no contribution from $\phi_{\mu\tau}$ at the leading order. 
 
	\item In the case of NO(upper panels), the variation of $P_{\mu e}$ with $\delta_{13}$ for 1100 km is sharper as 0.6 GeV is adjacent to the second oscillation maxima(0.7 GeV). However, in 295 km, the variation is less due to the first oscillation maxima occurring at 0.6 GeV. Thus probabilities at CP conserving values $0^\circ, \pm 180^\circ$ are more separated from probabilities at other CP violating values at 1100 km than at 295 km. 
	
	\item Also, in the case of NO, the maxima and minima of $P_{\mu e}$ occur at different $\delta_{13}$ values for 295 km and 1100 km. For instance, the probabilities at $\delta_{13}=\pm 90^\circ$ has a maximum difference from probabilities at CP conserving values for 295 km. However, in the case of 1100 km, the probabilities at $\delta_{13}=\pm 90^\circ$ are very close to probability values at $\pm 180^\circ$. Therefore, while evaluating the sensitivity to CP discovery at $\delta_{13}=\pm 90^\circ$, there will be a higher sensitivity for 295+1100 km configuration than individual 295 km and 1100 km due to the synergy.
	
	\item For IO, The variation with $\delta_{13}$ is very flat at 1100 km while the variation at 295 km remains similar to NO. This leads to poor sensitivity for CP discovery for the T2HKK configuration for IO.
\end{itemize}
The disappearance probability $P_{\mu\mu}$ doesn't depend on the CP phase at the leading order. Therefore, in the case of $P_{\mu\mu}$, dependence on $\phi_{\mu\tau}$ isn't linked with $\delta_{13}$.
\begin{figure}[!h]
	\centering
        \includegraphics[width=0.8\linewidth]{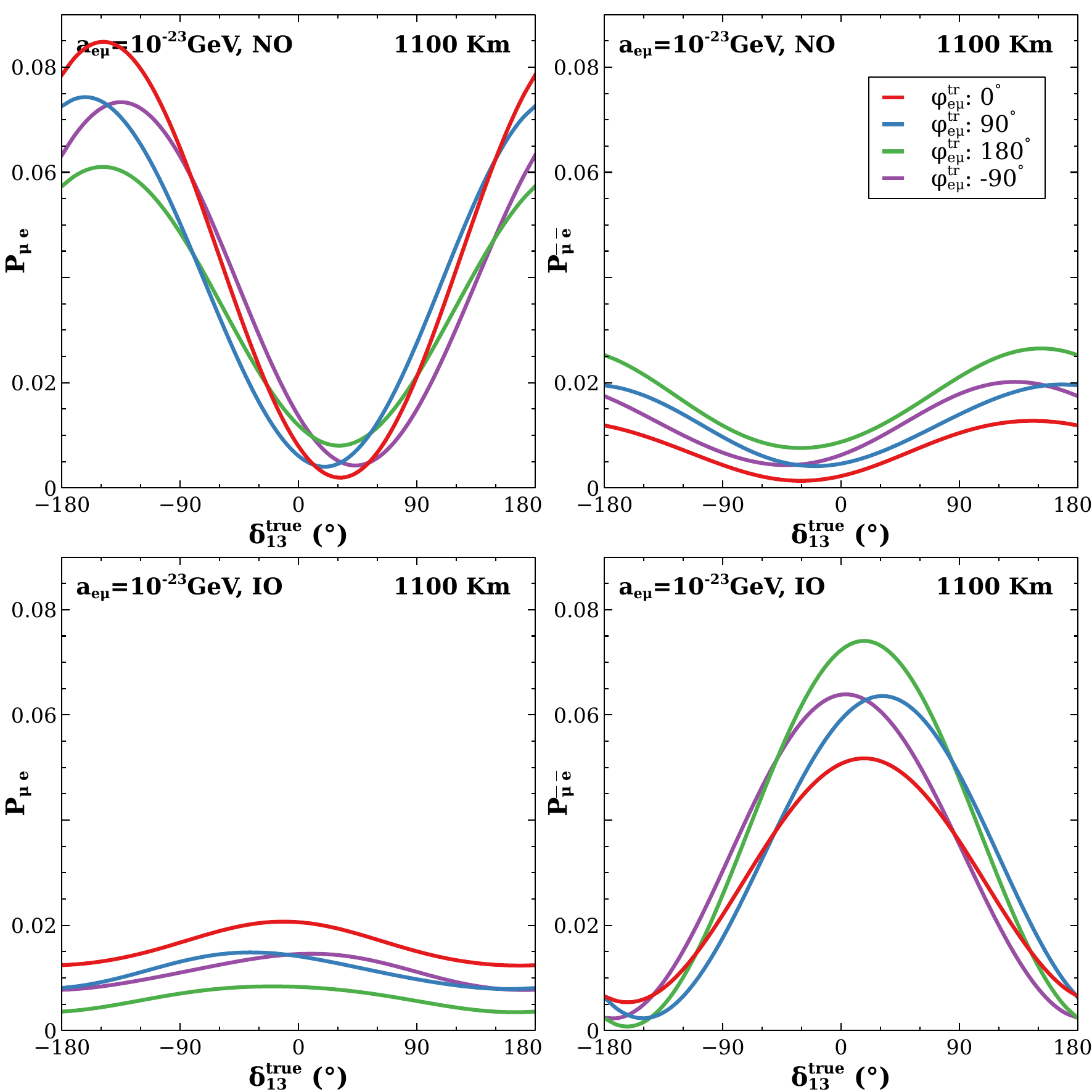}
	\includegraphics[width=0.8\linewidth]{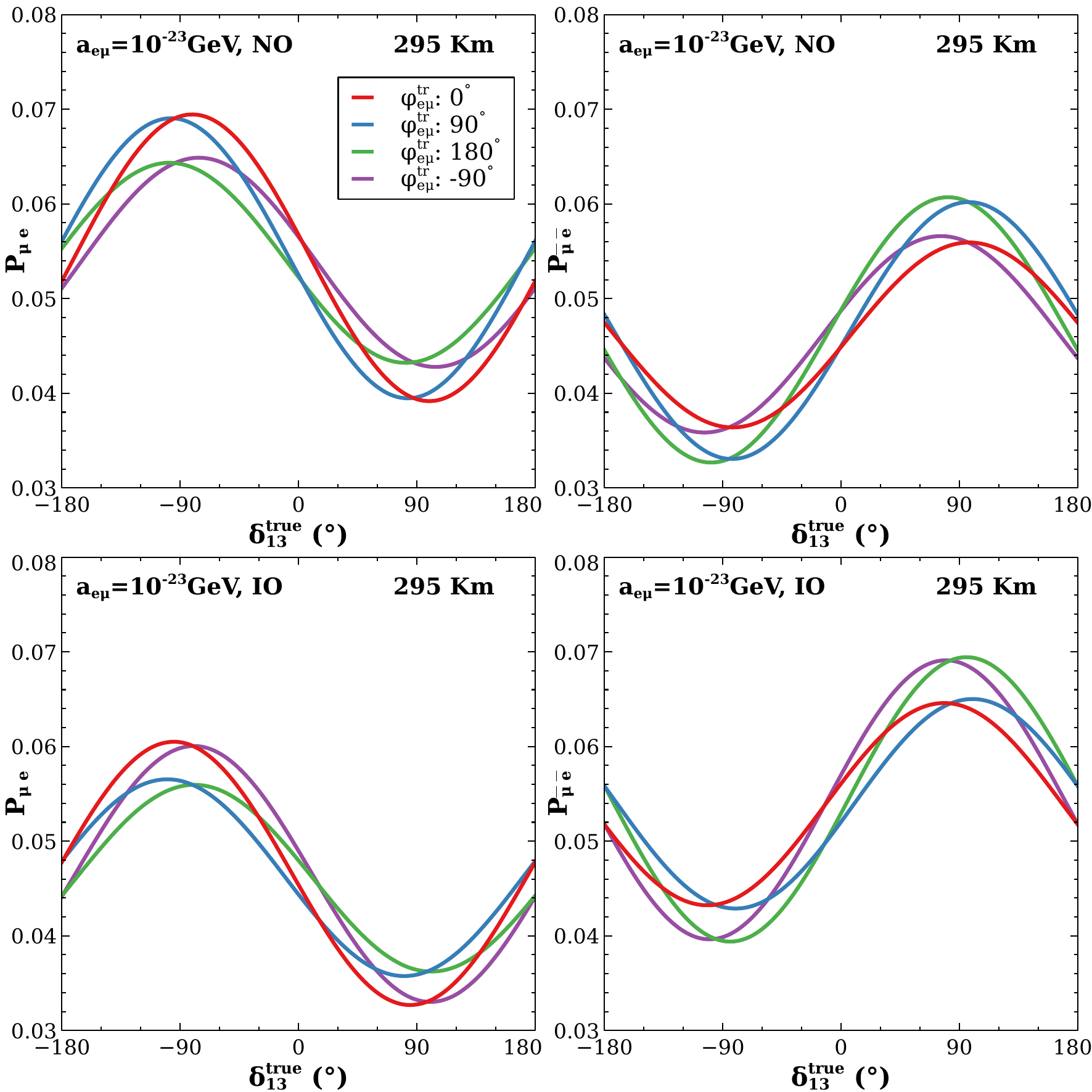}
	\caption{$ P_{\mu e} $ (left) , $P_{\bar{\mu}\bar{e}}$ (right) as a function of $\delta_{13}^{true}$ for true values of $\theta_{23}=49^\circ$, $ a_{e \mu}=10^{-23} $ GeV. The panels on the top (bottom) two rows refer to 295 km (1100 km) for NO(top) and IO(bottom). Violet, red, green, blue refer to $\phi_{e\mu}^{true}=-90^\circ,0^\circ,90^\circ,180^\circ$ respectively.}
	\label{fig:pme_dcp-aem}
\end{figure}

In fig.~\ref{fig:pme_dcp-aem}, the oscillation probabilities $P_{\mu e}, P_{\bar{\mu}\bar{e}}$ are plotted as a function of $\delta_{13}$ at fixed energy of 0.6 GeV corresponding to 295 km and 1100 km baselines for NO, IO considering $\theta_{23}=49^\circ$ and $a_{e\mu}=10^{-23}$ GeV. We observe the following features,

\begin{itemize}
    \item In 1100 km, the $P_{\mu e}$ probabilities have larger values than $P_{\bar{\mu}\bar{e}}$. However, in IO that order reverses.

    \item In NO, the $P_{\mu e}$ curves show a peak at the lower half plane (LHP)[$-180^\circ:0^\circ$] in the range $-160^\circ:-130^\circ$. The peaks of $P_{\bar{\mu}\bar{e}}$ curves occur at the upper half plane (UHP)[$0^\circ:180^\circ$] in the range $130^\circ:170^\circ$. 

    \item In the case of IO for 1100 km, both the $P_{\mu e}$, $P_{\bar{\mu}\bar{e}}$ has maxima around $0^\circ$.


    \item In 295 km, various probabilities for different values of $\phi_{e\mu}$ vary over a small region while being very close to each other.

    \item In 295 km, the maxima of $P_{\mu e}$, $P_{\bar{\mu}\bar{e}}$ curves occur around $\pm 90^\circ$ for both NO and IO. 

    \item In the case of both 295 km and 1100 km, the red (green) curves corresponding to $\phi_{e\mu}^{tr}=0^\circ(180^\circ)$ give the maximum (minimum) variation in $P_{\mu e}$. In $ P_{\bar{\mu}\bar{e}}$, this order reveres.
\end{itemize}

\section{\label{sec:chi}$ \chi^2 $ Analysis of CP discovery}
In this section, we study the potential of the experiments T2HKK/T2HK for CP discovery. The configurations for the proposed experiments are, (i) T2HK (Tokai to Hyper-Kamiokande): two detectors of 187 kton at 295 km, (ii)T2HKK (Tokai to Hyper-Kamiokande and Korea): one detector of 187 kton at 295 km and another similar detector at 1100 km away in Korea\cite{Hagiwara:2006vn}. For our study, we consider the second detector of T2HKK at an off-axis angle of $1.5^\circ$ from the source at J-PARC facility in Tokai\cite{Hyper-Kamiokande:2016srs}. T2HKK offers us the advantage of a larger matter effect at 1100 km than T2HK. For our numerical analysis with GLoBES\cite{Huber:2004ka, Huber:2007ji}, we use a proposed beam of energy 1.3 MW considering 2.5 years of neutrino mode and 7.5 years of anti-neutrino mode run time with an exposure of $27\times 10^{21}$ proton on target (POT). The detector configurations and systematic errors are taken from \cite{Hyper-Kamiokande:2016srs} and are tabulated in table~\ref{table:uncertainty}.

The final value of $\chi^2$ is derived after marginalization over pull variables $\xi$, and variables of oscillation $\omega$ as follows,
\begin{equation}\label{eq:chi-tot}
    \Delta \chi^2 = {\rm Min} [\chi^2_{stat}(\omega,\xi) + \chi^2_{pull}(\xi)],
\end{equation}
where $\chi^2_{pull}$ includes the symmetric errors and the Poissonian $\chi^2_{stat}$ is defined in terms of total true no of events $N_{i}^{true}$ and events generated by theoretical model $N_{i}^{test}$ in the $i^{th}$ energy beam.
\begin{gather}
    \chi^2_{stat}(\omega,\xi) = 2\sum_i [N_{i}^{test}-N_{i}^{true} + N_{i}^{true} \text{ln} \frac{N_{i}^{true}}{N_{i}^{test}}]\\
    \chi^2_{pull}=\sum_{r=1}^{4} \xi_r^2 
\end{gather}
The systematic uncertainties are included through the method of pull in terms of the variables: signal normalization error, background normalization error, energy calibration error on signal, and background (tilt). In this work the test parameters $\omega$ are $\theta_{23},\delta_{13},|\Delta_{31}|, a_{\alpha\beta},\phi_{\alpha\beta}$.
    \begin{table}[H]
	\centering
	\begin{tabular}{|c|c|c|}
		\hline
		Parameter & True Value & Marginalization Range\\\hline
		$\theta_{12}$ & $33.4^\circ$  & N.A.\\
		$\theta_{13}$ & $8.62^\circ$ & N.A.\\
		$\theta_{23}$ & $49^\circ$ & $39^\circ:51^\circ$\\
		$\delta_{13}$ & $-180^\circ:180^\circ$& $0^\circ,180^\circ$ \\
		$|\Delta_{31}|$ & $2.5\times 10^{-3}$ eV$^2$ & $2.4:2.6\times 10^{-3}$ eV$^2$\\
		$a_{\alpha\beta}$ & $10^{-23}$ GeV & $10^{-22}:10^{-24}$ GeV\\
		$\phi_{\alpha\beta}$ & $-180^\circ:180^\circ$& $0^\circ,180^\circ$ \\
		\hline
	\end{tabular}\label{table:chi-parameter}
        \caption{True values of all the parameters and their range of marginalization}
    \end{table}
    \begin{table}[H]
        \centering
		\begin{tabular}{|c|c|c|}
			\hline
			Channel & 295 km & 1100 km\\\hline
			$\nu_e$ Appearance  &\hspace{2mm}$3.2\%(5\%)$ &\hspace{2mm} $3.8\%(5\%)$\\
			$\nu_\mu$ Disappearance  &\hspace{2mm}$3.6\%(5\%)$ &\hspace{2mm} $3.8\%(5\%)$\\
			$\bar{\nu}_e$ Appearance &\hspace{2mm}$3.9\%(5\%)$ &\hspace{2mm} $4.1\%(5\%)$\\
                $\bar{\nu}_\mu$ Disappearance &\hspace{2mm}$3.6\%(5\%)$ &\hspace{2mm} $3.8\%(5\%)$\\
		    \hline
		\end{tabular}\label{table:uncertainty}
        \caption{The signal (background) normalization uncertainties of the experiments for different channels}
    \end{table}
We have seen in table \ref{table:nsi-bound} the current bound for NSI parameters are $\sim 10^{-23}$ GeV. Therefore, we have considered true values of $a_{e\mu}, a_{e\tau}, a_{\mu \tau}=10^{-23}$ GeV throughout our study. For numerical analysis for CP discovery, the test values are considered as $\delta_{13}=0^\circ,180^\circ$. Similarly, in the presence of an extra LIV phase, we consider test values of $\delta_{13},\phi_{\alpha\beta}$ as combinations of $0^\circ,180^\circ$. While performing chi-square($\chi^2$) analysis in the presence of LIV, we consider one parameter to be non-zero at a time. Apart from phases, we have marginalized the chi-square over $\theta_{23}$ and $|\Delta_{31}|$. The true values\cite{Esteban:2020cvm} and the marginalization ranges of the parameters are given in table \ref{table:chi-parameter}. The run time in neutrino and anti-neutrino mode is 2.5 years and 7.5 years, respectively.

\subsection{Single detector analysis}
In this section, the sensitivity to CP discovery is probed with a single detector at 295 km and 1100 km. This helps in understanding the features of these individual baselines. The total event rates get equal contributions from neutrinos and anti-neutrinos because of the chosen run time. Therefore studying sensitivity for individual channels will help in understanding the total sensitivity. In this section, we study the effect of only the LIV parameter $a_{e\mu}$ as a representative case.
\begin{figure}[H]
	\centering
	\includegraphics[width=0.95\linewidth]{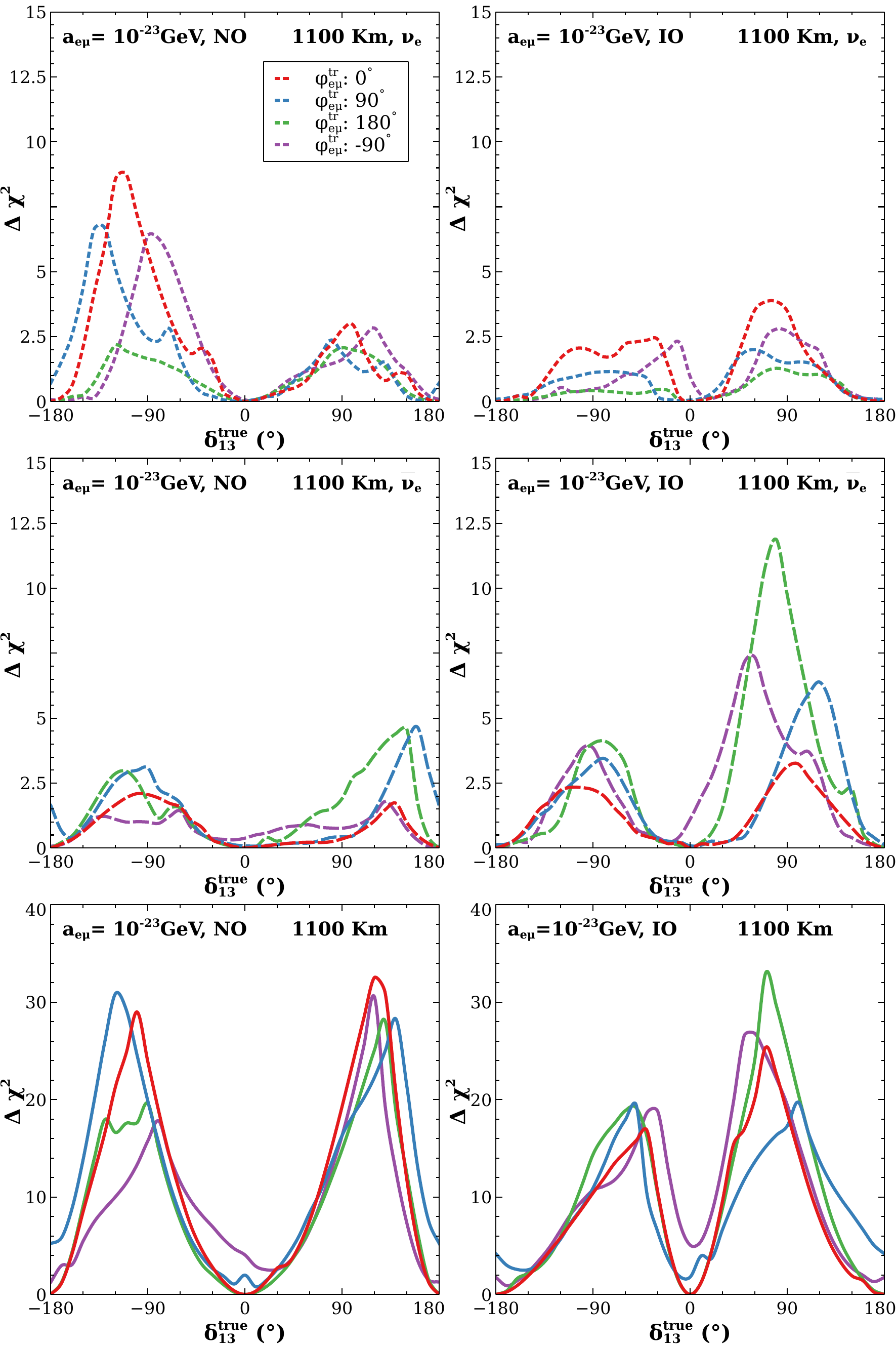}
	\caption{$ \chi^2 $ in $\nu_e$ (top), and $\bar{\nu}_e$ (middle) channel and  total $\chi^2$ (bottom) as a function of $\delta_{13}^{true}$ for true values of $\theta_{23}=49^\circ$ with $ a_{e\mu}=10^{-23} $ GeV at 1100 km for NO (left), IO(right). Violet, red, green blue refer to $\phi_{e\mu}=-90^\circ,0^\circ,90^\circ,180^\circ$ respectively.}
	\label{fig:chi-aem-e,eb-1100}
\end{figure}
The $\chi^2$ for appearance channel in neutrino, anti-neutrino mode, along with the total $\chi^2$ for 1100 km is plotted in the top, middle, and bottom panels of fig.~\ref{fig:chi-aem-e,eb-1100}. The left(right) panels correspond to the NO (IO). The different true values of $\phi_{e\mu}=-90^\circ,0^\circ,90^\circ,180^\circ$ have been shown by violet, red, blue and green curves, respectively. The features of significance in fig~ \ref{fig:chi-aem-e,eb-1100} are as follows,
\begin{itemize}
    \item In the neutrino mode (NO), the red curve $\phi_{e\mu}=0^\circ$ has the maximum sensitivity in both half-planes, but in LHP, the magnitude at the peak is significantly larger. The green curve $\phi_{e\mu}=180^\circ$ has the lowest sensitivity. This is consistent with the features seen from the plot of $P_{\mu e}$ in the left panels of fig.~\ref{fig:pme_dcp-aem}. 

    \item In the anti-neutrino mode (NO),  the highest sensitivity is achieved for the green $\phi_{e\mu}=-90^\circ$ and blue curve $\phi_{e\mu}=180^\circ$ in both UHP and LHP but the peak value in UHP is higher. 

    \item In the case of total sensitivity (NO), the red curve ($\phi_{e\mu}=0^\circ$) has the highest sensitivity in the UHP, and the blue curve ($\phi_{e\mu}=180^\circ$) reach the maximum sensitivity in the LHP. While marginalizing, the minimum of $\chi^2$ occurs at different values of the parameters for neutrino and anti-neutrino, leading to a synergistic effect in total sensitivity.

    \item In the case of neutrino mode for IO, the red (green) shows the maximum (minimum) sensitivity, and the value of $\chi^2$ is higher in UHP than LHP.

    \item In the context of anti-neutrino mode for IO, the green (red) curve reaches the maximum (minimum) value of $\chi^2$. The green curve's maximum value of $\chi^2$ is predominantly the highest in UHP. The other curves also have maxima of higher value in UHP.
    

     \item In the case of IO for the total sensitivity, the green curve ($\phi_{e\mu}=0^\circ$) has the maximum $\chi^2$ in both LHP and UHP with the latter case having significantly higher value.

     \item The sensitivity curves for $\phi_{e\mu}=-90^\circ, 90^\circ$ show non-zero sensitivity at $\delta_{13}^{true}=0^\circ,\pm 180^\circ$. This happens as the test values $\phi_{e\mu},\delta_{13}$ are a combination of $0^\circ,180^\circ$ that is different from these true sets of values.
\end{itemize}

In fig.~\ref{fig:chi-aem-e,eb-295}, the value of $\chi^2$ is depicted as a function of $\delta_{13}^{true}$ for a single detector at 295 km considering true ordering as NO (left) and IO (right). The top, middle, and bottom panels refer to the sensitivity in neutrino, anti-neutrino, and the total sensitivity, respectively. The different colors are for various $\phi_{e\mu}$ are similar to fig.~\ref{fig:chi-aem-e,eb-1100}. The main observations of fig.~\ref{fig:chi-aem-e,eb-295} are as follows,
\begin{itemize}
    \item In the case of both NO and IO, in the neutrino mode, the red ($\phi_{e\mu}=0^\circ$) and green ($\phi_{e\mu}=180^\circ$) curves show the maximum and minimum sensitivity respectively in LHP, but In UHP all curves have very low and similar sensitivity. In the antineutrino mode, the order of sensitivity for all the curves is almost similar, but the green curve ($\phi_{e\mu}^{tr}=180^\circ$) has slightly more sensitivity.


    \item In both NO and IO, we observe higher sensitivity in neutrino mode. This is due to the fact that $P_{\mu e}$ curves have a higher range of variation than $P_{\bar{\mu}\bar{e}}$ ones as was seen in fig.~\ref{fig:pme_dcp-aem}.

    \item In the case of total sensitivity, all the curves have similar sensitivity except for the red curve ($\phi_{e\mu}^{tr}=0^\circ$) showing a slightly higher value of $\chi^2$ in UHP. 
    
\end{itemize}
The synergy between various channels in the test $\theta_{23}$ is depicted in fig.~\ref{fig:chi-aem-syn} for 295 km (left panel) and 1100 km (right panel). We observe that the minimum of the total chi-square is obtained at one of the minima of the $\nu_\mu,\bar{\nu}_\mu$ channels around $\theta_{23}=49^\circ$. However, the total chi-square value is boosted by the non-zero chi-square contribution from the $\nu_e,\bar{\nu}_e$ channels.

\begin{figure}[H]
	\centering
	\includegraphics[width=0.95\linewidth]{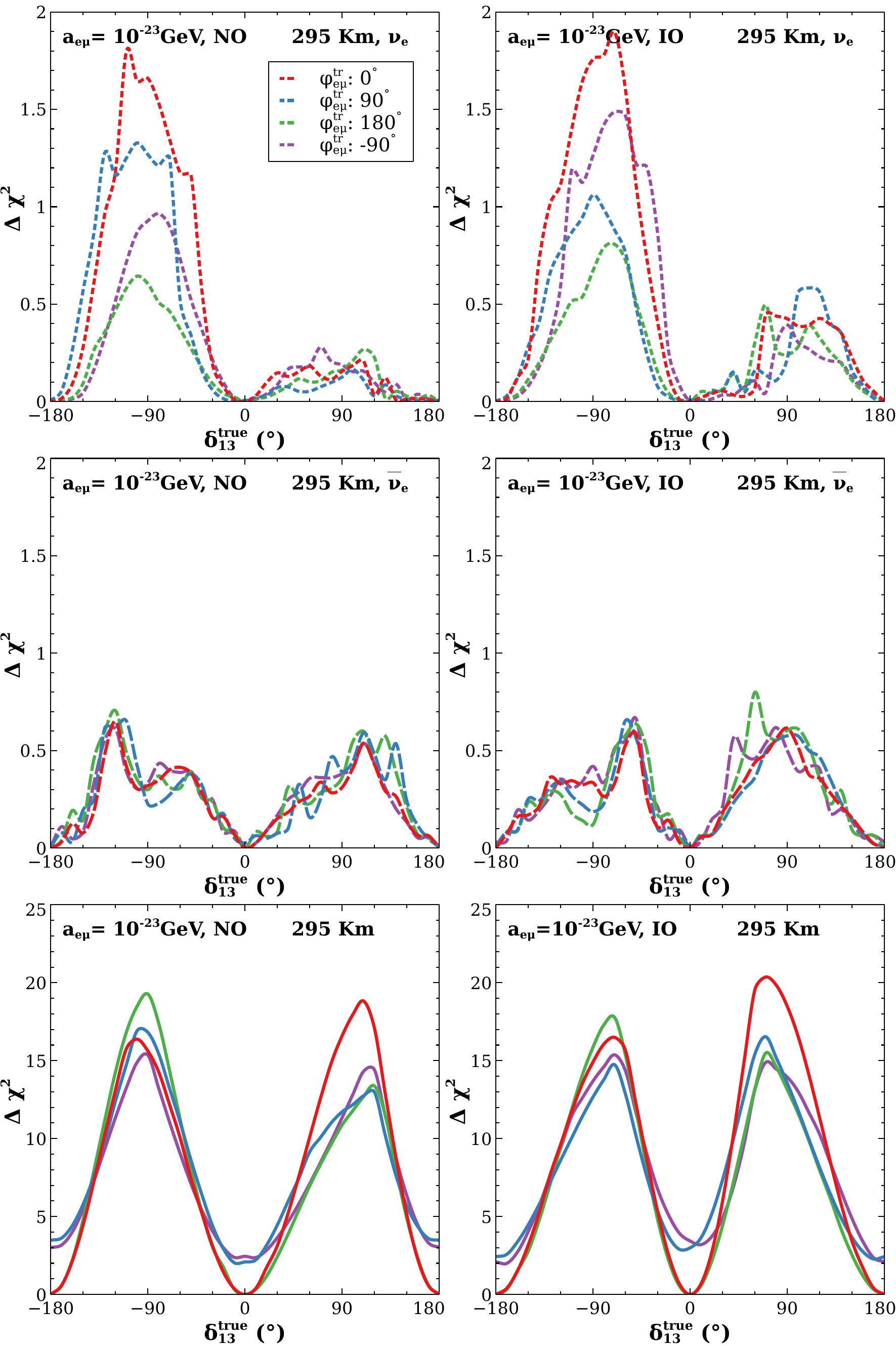}
	\caption{$ \chi^2 $ in $\nu_e$ (top), and $\bar{\nu}_e$ (middle) channel and  total $\chi^2$ (bottom) as a function of $\delta_{13}^{true}$ for true values of $\theta_{23}=49^\circ$ with $ a_{e\mu}=10^{-23} $ GeV at 1100 km for NO (left), IO(right). Violet, red, green blue refer to $\phi_{e\mu}=-90^\circ,0^\circ,90^\circ,180^\circ$ respectively.}
	\label{fig:chi-aem-e,eb-295}
\end{figure}
\begin{figure}[H]
    \centering
    \includegraphics[width=0.85\linewidth]{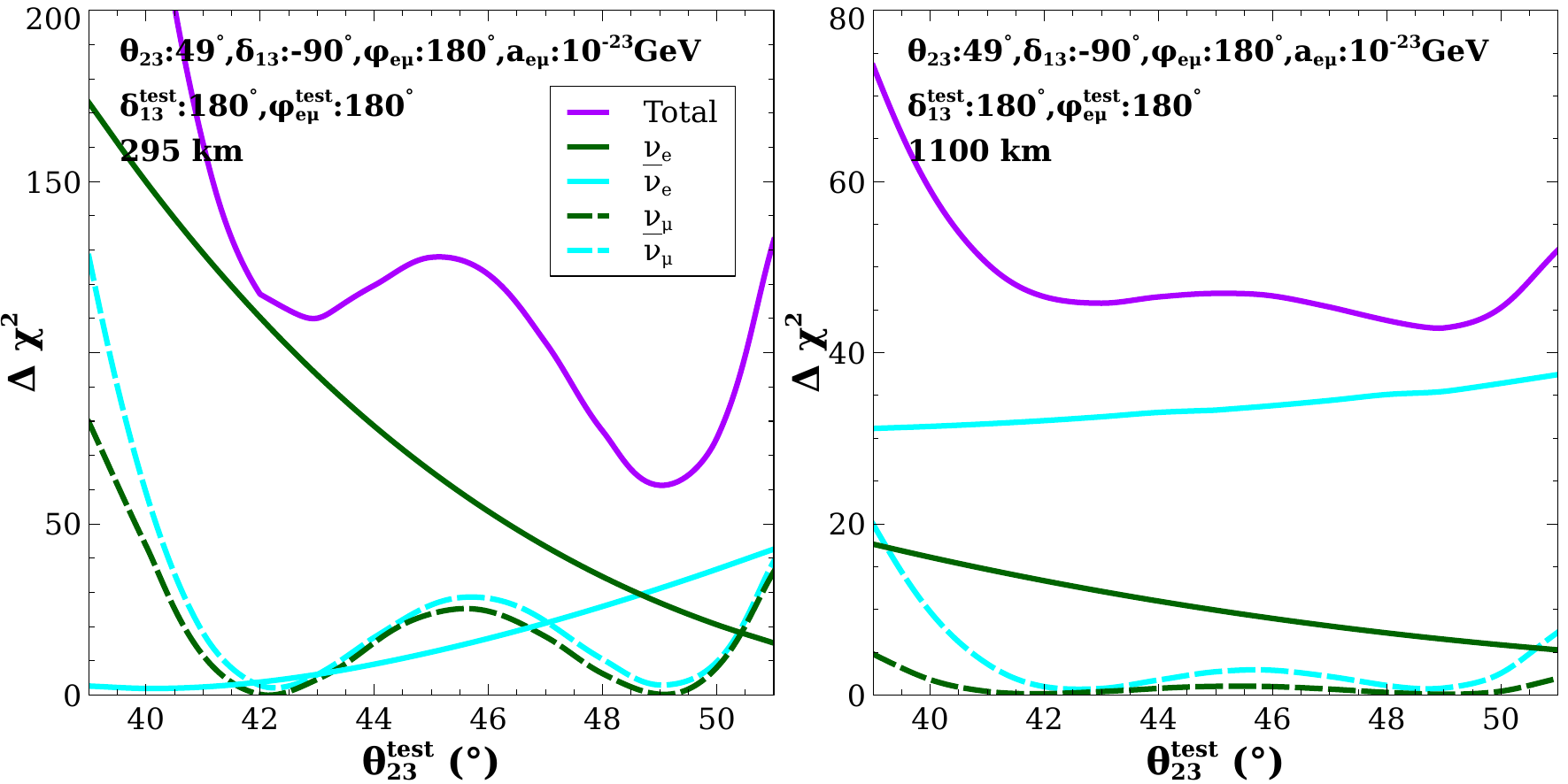}
    \caption{$\chi^2$ as a function of $\theta_{23}^{test}$ at 295 km (left), 1100 km (right).  Green (blue) refers to $\nu(\bar{\nu}$) channels and violet gives total $\chi^2$. }
    \label{fig:chi-aem-syn}
\end{figure}

\subsection{Comparative analysis between T2HKK and T2HK}
\begin{figure}[H]
	\centering
	\includegraphics[width=0.9\linewidth]{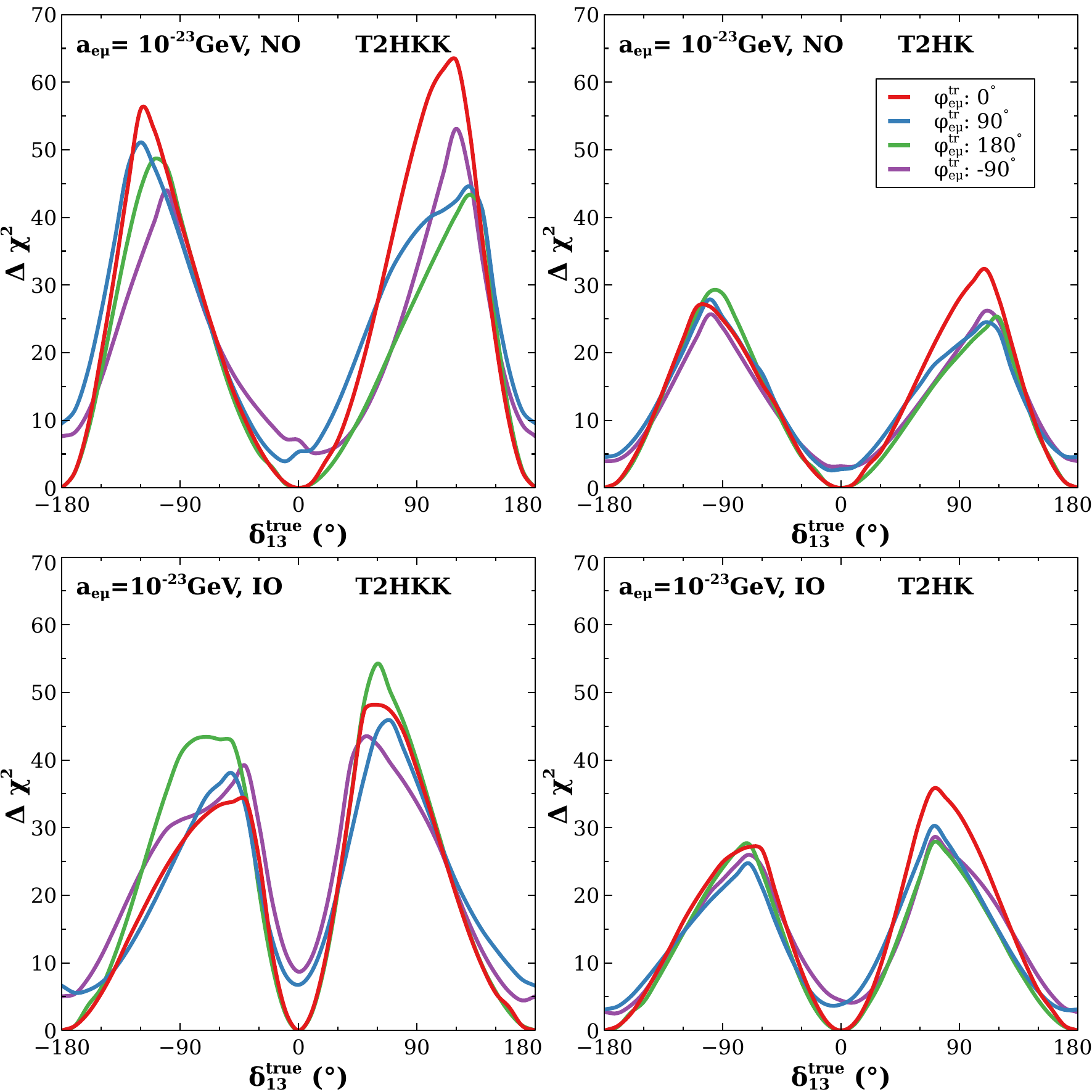}
	\caption{$ \chi^2 $ as a function of $\delta_{13}^{true}$ for true values of $\theta_{23}=49^\circ$, $ a_{e\mu}=10^{-23} $ GeV in T2HKK (left) and T2HK (right) configurations for NO (top), and IO(bottom). Violet, red, green, and blue curves refer to $\phi_{e\mu}^{true}=-90^\circ,0^\circ,90^\circ,180^\circ$ respectively.}
	\label{fig:chi-aem-t2hk}
\end{figure}

In this section, we compare and contrast the CP discovery potential of the proposed T2HKK and T2HK configurations. This study is performed for the LIV parameters $a_{e\mu},a_{e\tau},a_{\mu\tau}$ taking one of these to be non-zero at a time. In fig.~\ref{fig:chi-aem-t2hk}, we present the sensitivity as a function of $\delta_{13}^{true}$ for T2HKK (left) and T2HK (right) for NO (top) and IO(bottom) for $a_{e\mu}^{true}=10^{-23}$ GeV. Different curves correspond to the different values of $\phi_{e\mu}^{true}$.

\begin{figure}[H]
	\centering
	\includegraphics[width=0.9\linewidth]{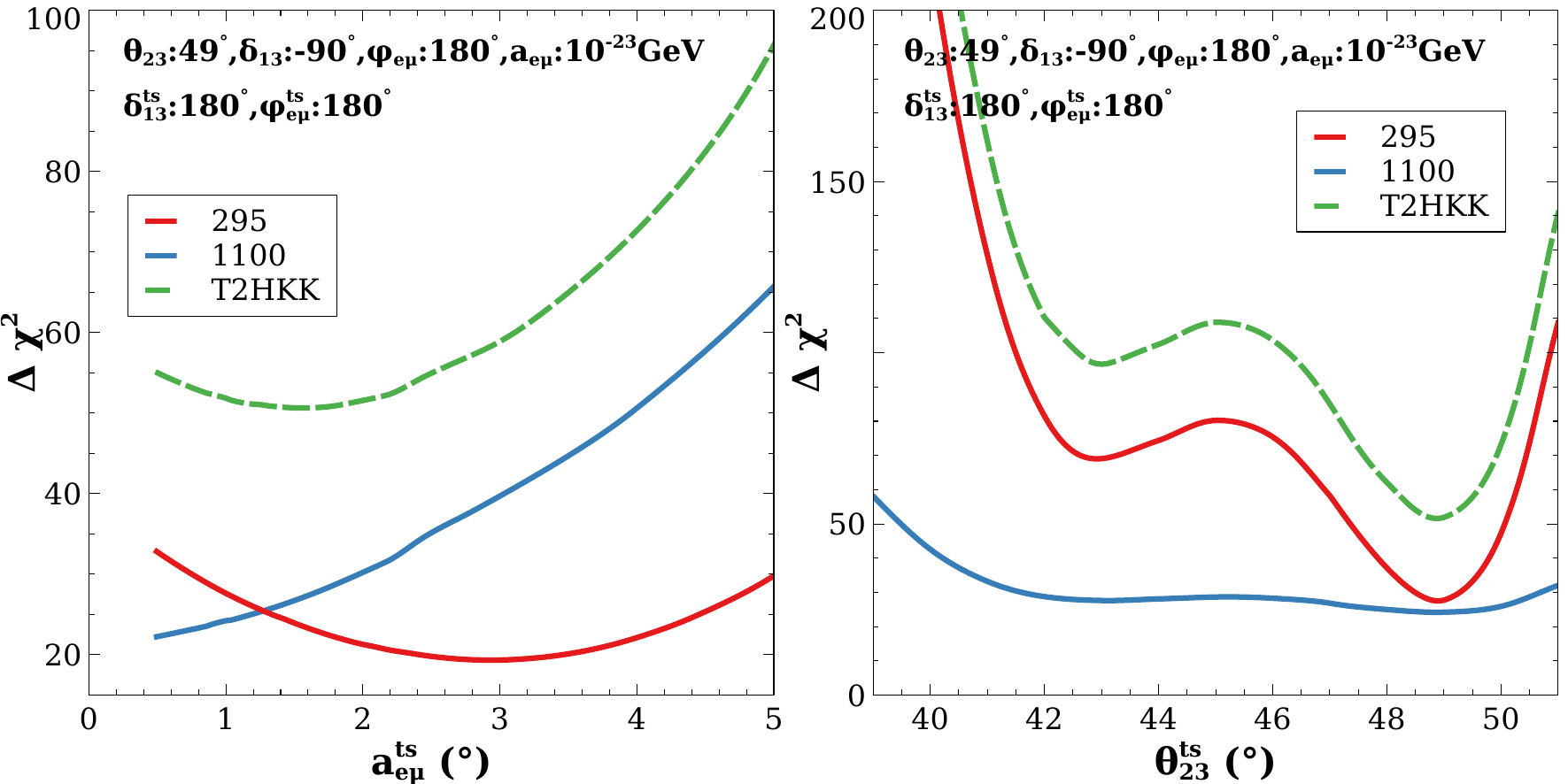}	
	\caption{$ \chi^2 $ as a function of $a_{e \mu}^{test}$(left), $ \theta_{23}^{test} $(right) for true values of $\theta_{23}=49^\circ$, $\delta_{13}=-90^\circ$, $ \phi_{e\mu}=180^\circ $, $ a_{e \mu}=10^{-23}$GeV.}
	\label{fig:chi-em-test}
\end{figure}

The major points observed from fig.~\ref{fig:chi-aem-t2hk} are as follows; 
\begin{itemize}
    \item T2HKK offers the best sensitivity for all values of the LIV phase $\phi_{e\mu}$. This is due to the synergistic effect between 1100 km and 295 km baselines. This will be explained later in the context of fig.~\ref{fig:chi-em-test}.

    \item The highest sensitivity is obtained at $\delta_{13}=90^\circ$ for both T2HK and T2HKK. The corresponding values of $\phi_{e\mu}$ are $0^\circ$($180^\circ$) for NO(IO) case in T2HKK, and $0^\circ$ in T2HK. This can be understood from fig. 3 and 4, which shows that for individual baseline, the maxima comes at $\phi_{e\mu}=0^\circ$ around $\delta_{13}=90^\circ$.

    

    
\end{itemize}

\begin{figure}[H]
	\centering
	\includegraphics[width=0.9\linewidth]{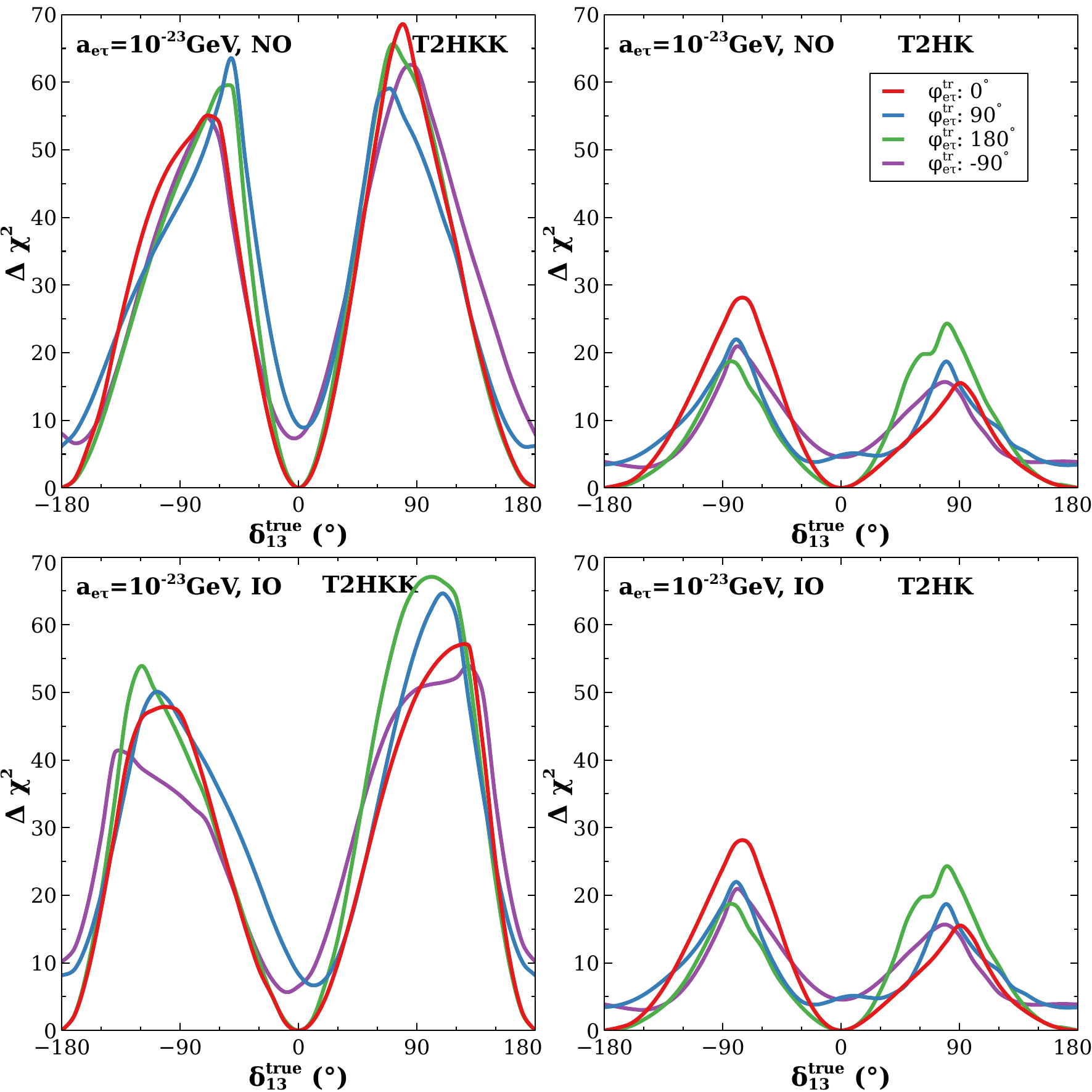}
	\caption{$ \chi^2 $ as a function of $\delta_{13}^{true}$ for true values of $\theta_{23}=49^\circ$ with $ a_{e\tau}=10^{-23} $ GeV for T2HKK (left) and T2HK (right) configurations in NO (top), and IO (bottom). Violet, red, green, and blue curves refer to $\phi_{e\mu}^{true}=-90^\circ, 0^\circ, 90^\circ, 180^\circ$ respectively.}
	\label{fig:chi-aet}
\end{figure}

In order to understand the synergy between 295km and 1100 km baselines, in fig.~\ref{fig:chi-em-test}, we have shown the $\chi^2$ as a function of test $a_{e\mu}$ (left), and $\theta_{23}$ (right) for a set of true parameters keeping other test parameters fixed.
\begin{itemize}
    \item From the left panel, we see that the minimum $\chi^2$ for 295 km and 1100 km occurs at different test values of LIV parameter $ a_{e\mu}$. Whereas in the case of T2HKK, the minimum occurs at a different test value of $a_{e\mu}$, thus enhancing the $\Delta \chi^2$ since both the baselines contribute.
	
    \item In the right panel, the enhancement in $\chi^2$ for T2HKK is due to the increased statistics. However, when marginalizing the $\chi^2$ over other test parameters, the synergy is also observed in $\theta_{23}$.
 
\end{itemize}
In fig. \ref{fig:chi-aet}, we present the values of $\chi^2$ as a function of $\delta_{13}^{true}$ for $a_{e\tau}^{true}=10^{-23}$ GeV in the T2HKK and T2HK configurations corresponding to NO (IO) in the top (bottom) column. The results for the true values of phase $\phi_{e\tau}$ as $-90^\circ$(violet), $0^\circ$(red), $90^\circ$(blue), $180^\circ$(green) using different colours as mentioned in the parenthesis. The major observations are as follows,
\begin{itemize}
    \item Similar to in fig.~\ref{fig:chi-aem-t2hk}, the sensitivity at T2HKK is quiet higher than T2HK configurations.

    \item We observe the maximum sensitivity in T2HKK around $\delta_{13}= 90^\circ$ ($-90^\circ$) which is influenced by the maxima of $P_{\mu e}$ ($P_{\bar{\mu}\bar{e}}$) curves in 295 km occurring at $90^\circ$ ($-90^\circ$). Although most of the curves show sensitivity in a similar range, the red ($\phi_{e\tau}=0^\circ$) one reaches the highest at UHP of $\delta_{13}^{true}$.
\end{itemize}

\begin{figure}[H]
	\centering
	\includegraphics[width=0.9\linewidth]{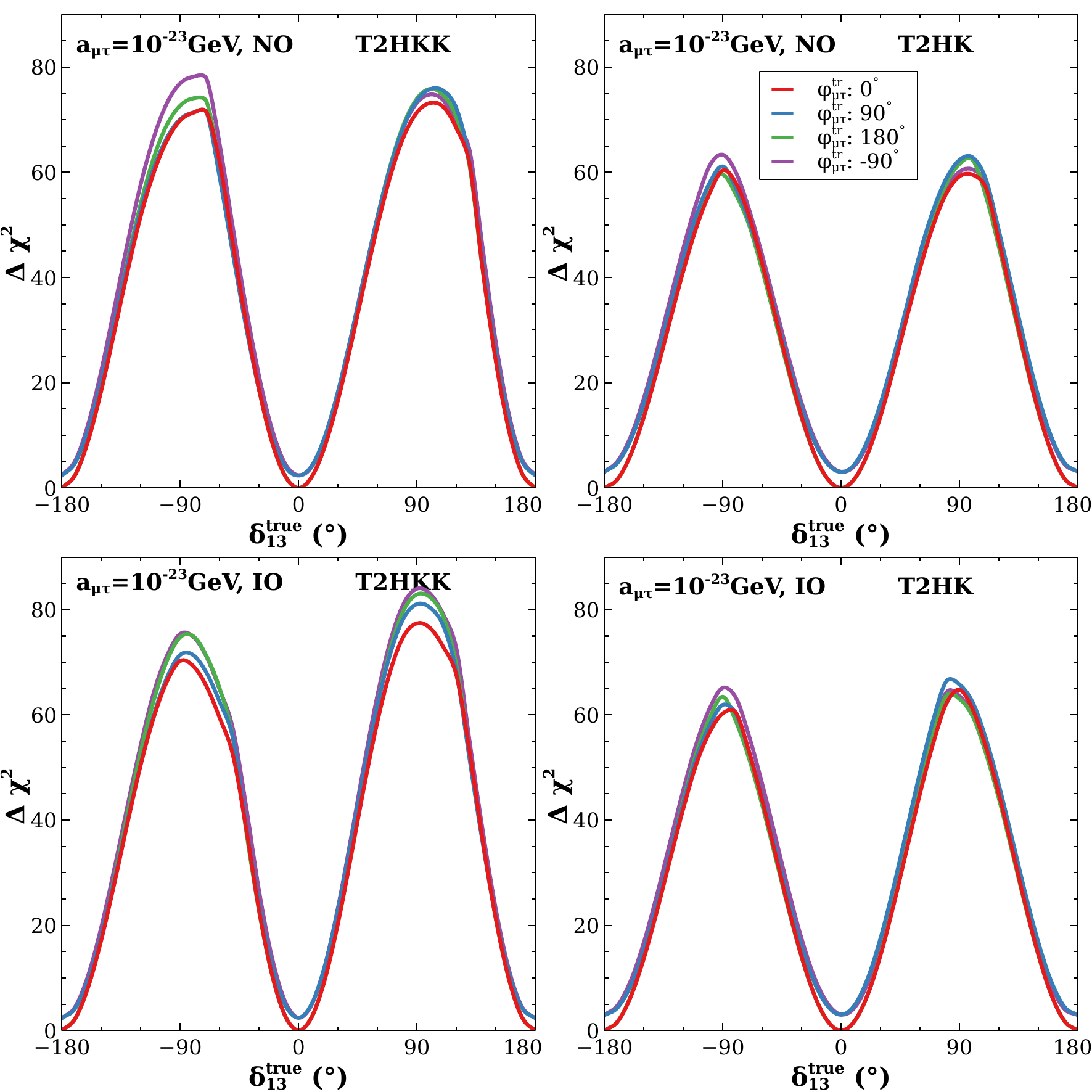}
	\caption{$ \chi^2 $ as a function of $\delta_{13}^{true}$ for true values of $\theta_{23}=49^\circ$ with $ a_{\mu\tau}=10^{-23}$ GeV for T2HKK (left) and T2HK (right) configurations for NO (top), and IO (bottom). Violet, red, green and blue curves refer to $\phi_{\mu\tau}^{true}=-90^\circ,0^\circ,90^\circ,180^\circ$ respectively.}
	\label{fig:chi-amt}
\end{figure}
We show the $\chi^2$ as a function of true $\delta_{13}$ in the fig.~\ref{fig:chi-amt} for effects of $a_{\mu\tau}$ for T2HKK and T2HK configurations in NO(top) and IO(bottom). The noteworthy points from these two figures are as follows,
\begin{itemize}
	\item The best sensitivity is observed in T2HKK, but the sensitivity of T2HK is also very close. The reason behind this is no significant effect of $a_{\mu\tau}$ in $P_{\mu e}$.
	
	\item Also, there is no significant variation of sensitivity w.r.t phase $\phi_{\mu\tau}$. This is due to the narrow band of due to $\phi_{\mu\tau}$ as also seen from probability plots in fig.~\ref{fig:pme-dcp}.
\end{itemize}

\section{Precision $\chi^2$ analysis of $\delta_{13},\phi_{\alpha\beta}$'s}\label{sec:precision-LIV}
\begin{figure}[H]
	\centering
		\includegraphics[width=0.96\linewidth]{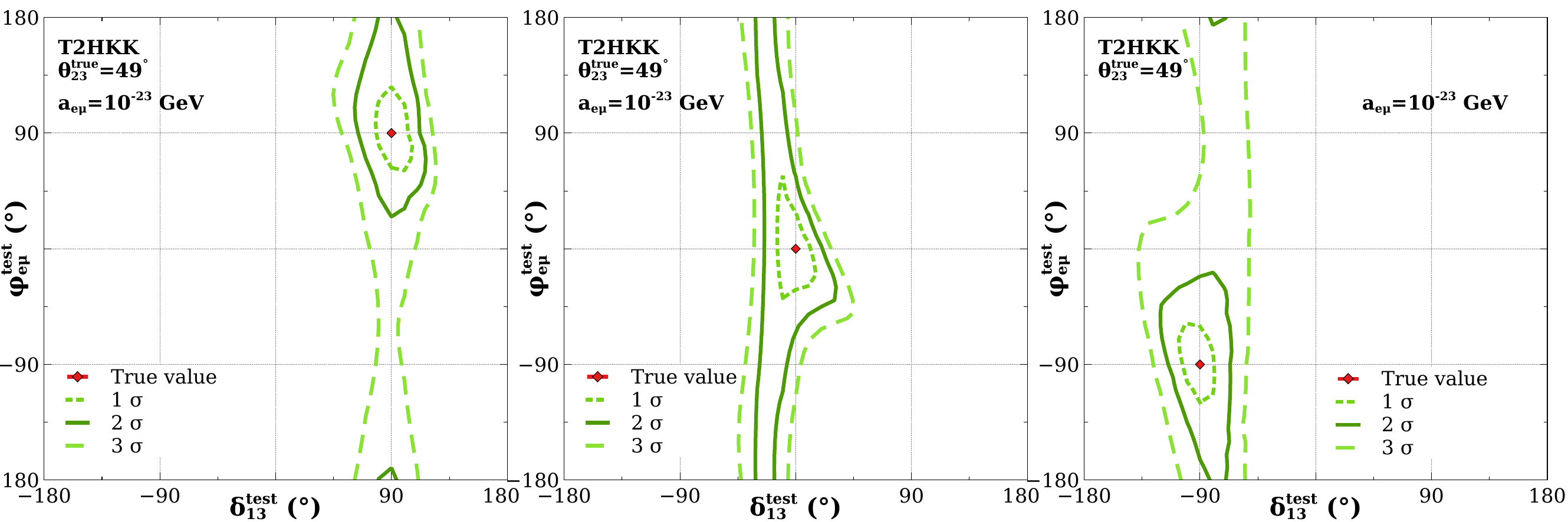}
		\includegraphics[width=0.96\linewidth]{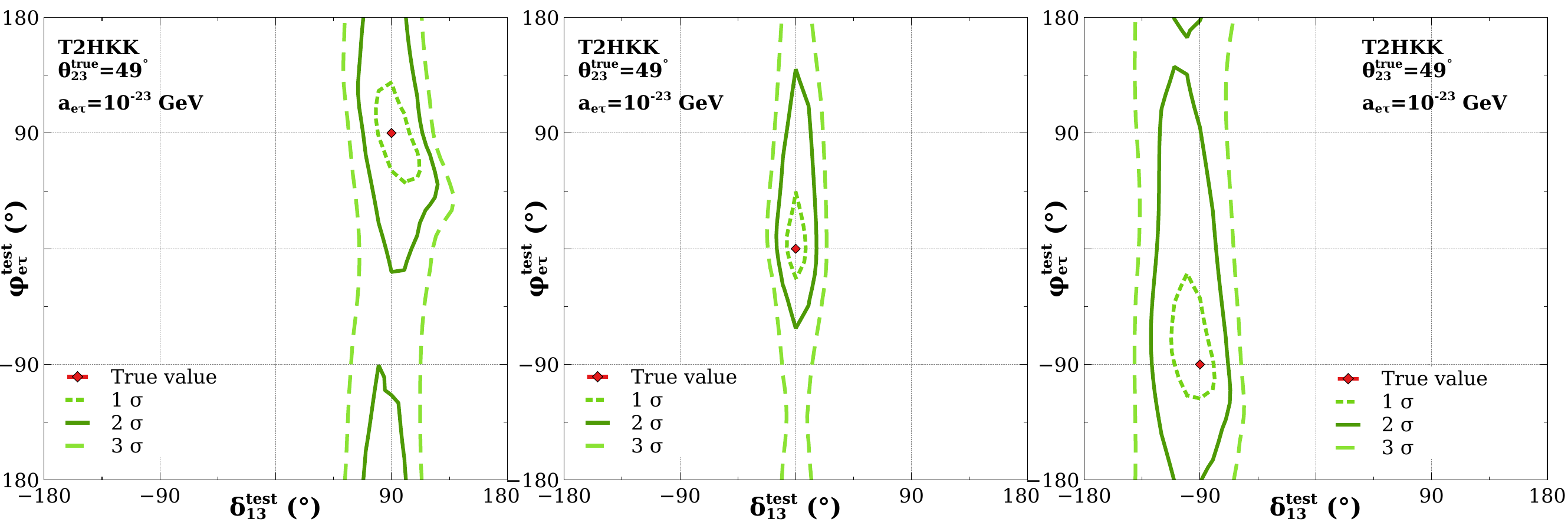}
		\includegraphics[width=0.96\linewidth]{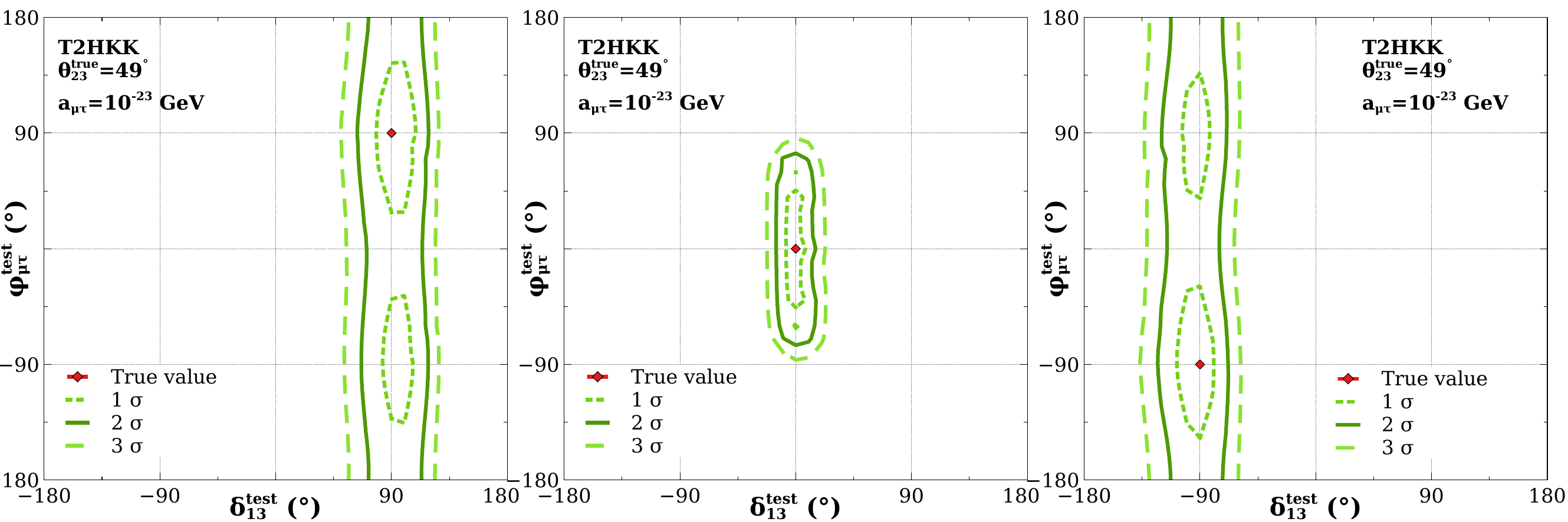}
		\caption{$1\sigma$(dotted), $2\sigma$(solid), $3\sigma$(dashed) contours[2 d.o.f.] corresponding to three different true values of $\delta_{13},\phi_{jk}$ for true LIV parameters $a_{e\mu}$ (top), $a_{e\tau}$ (middle) and $a_{\mu\tau}$ (bottom) having value of $10^{-23}$ GeV for T2HKK configuration}
		\label{fig:chi-pre-t2hkk}
    \end{figure}

\begin{figure}[H]
		\centering
		\includegraphics[width=0.96\linewidth]{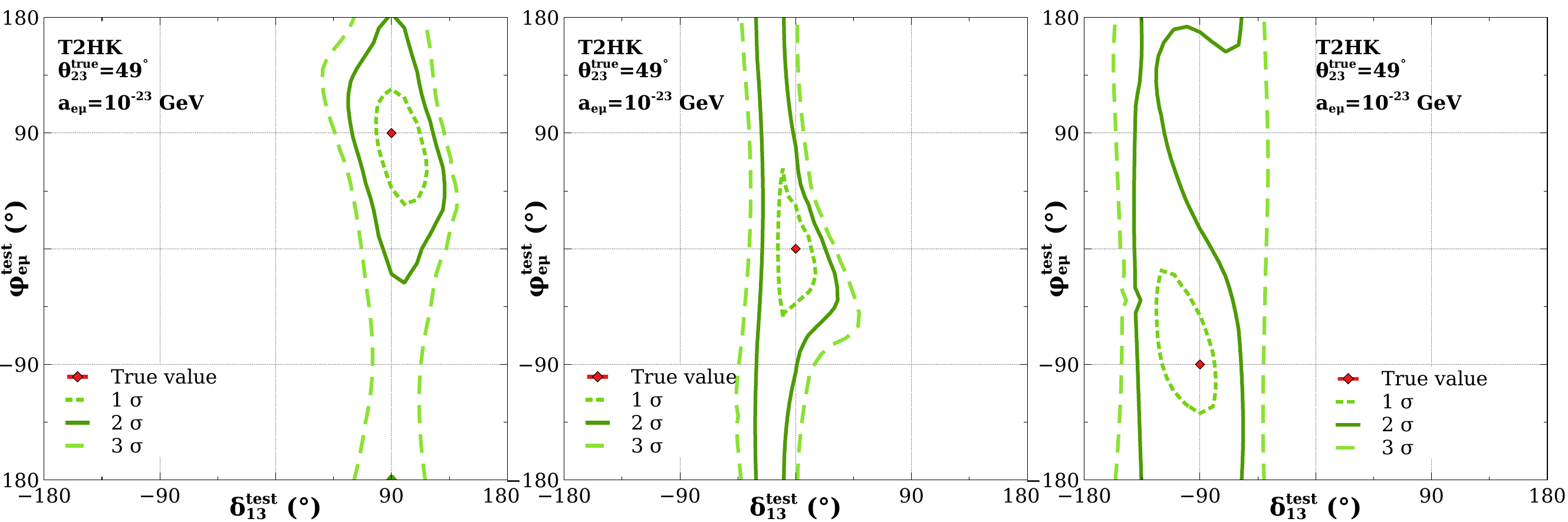}
		\includegraphics[width=0.96\linewidth]{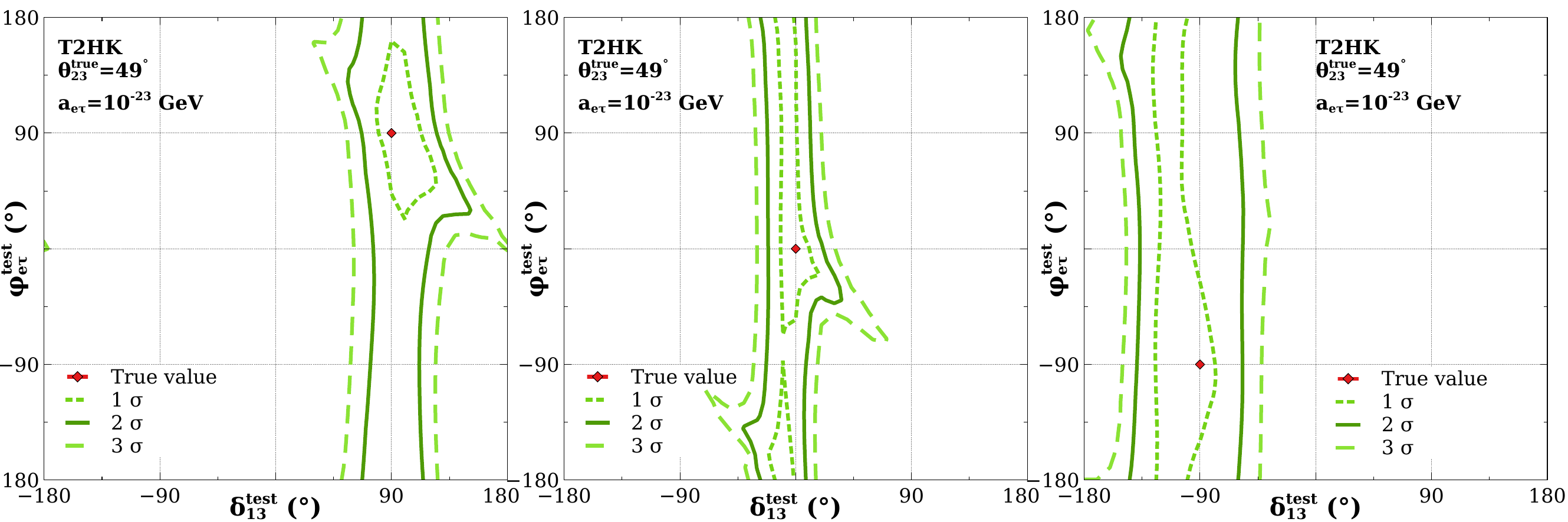}
		\includegraphics[width=0.96\linewidth]{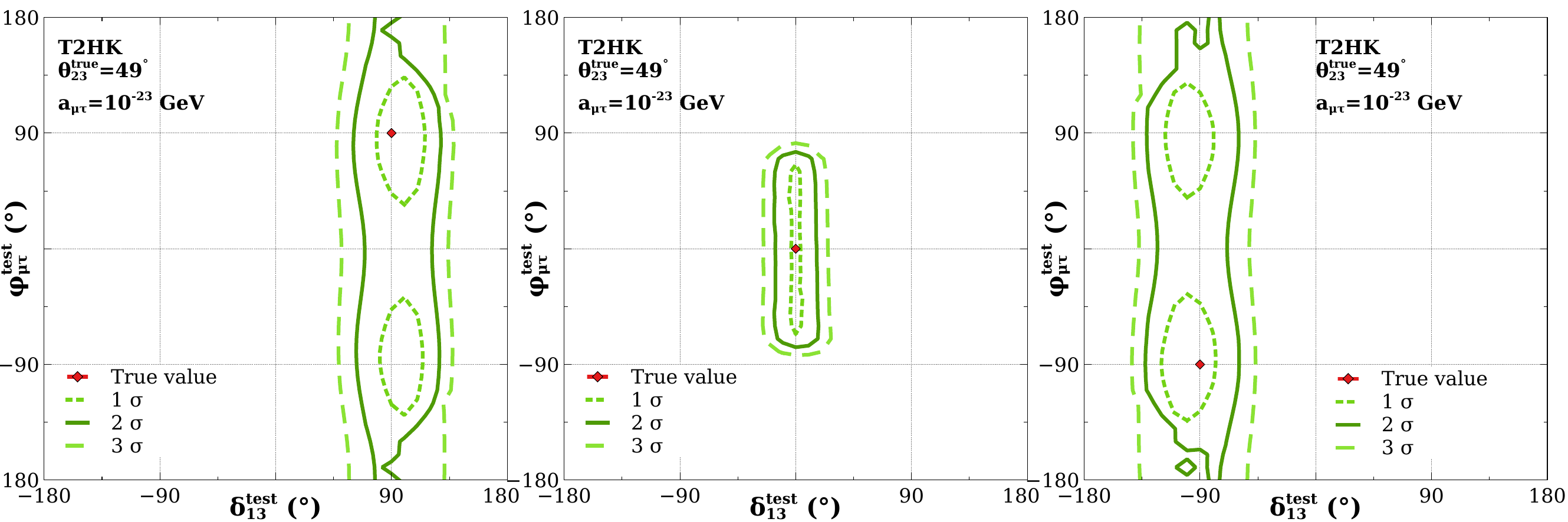}
		\caption{$1\sigma$(dotted), $2\sigma$(solid), $3\sigma$(dashed) contours[2 d.o.f.] corresponding to three different true values of $\delta_{13},\phi_{jk}$ for true LIV parameters $a_{e\mu}$ (top), $a_{e\tau}$ (middle) and $a_{\mu\tau}$ (bottom) having value of $10^{-23}$ GeV for T2HK configuration}
		\label{fig:chi-pre-t2hk}
    \end{figure}

In this section, we present the precision of $\delta_{13}$ and LIV phases $\phi_{\alpha\beta}$'s for T2HKK, T2HK in the figures \ref{fig:chi-pre-t2hkk} and \ref{fig:chi-pre-t2hk} respectively. These are presented in terms of contours in $\phi_{\alpha\beta}-\delta_{13}$ test plane of various combinations of true values of $\phi_{\alpha\beta},\delta_{CP}=0^\circ, 90^\circ, -90^\circ$. We consider the true values of LIV parameters as $10^{-23}$ GeV, $\theta_{23}=49^\circ$.
we can observe the following points from fig.~\ref{fig:chi-pre-t2hkk},
\begin{itemize}
    \item In the topmost panels corresponding to $a_{e\mu}$, we observe closed $2\sigma$ contours for $\delta_{13},\phi_{e\mu}=90^\circ,-90^\circ$ but not for $\delta_{13},\phi_{e\mu}=0^\circ$(middle panel).

    \item On the other hand, in the middle panels corresponding to $a_{e\tau}$, the $2\sigma$ precision is better for $\delta_{13},\phi_{e\tau}=0^\circ$  but worse for $90^\circ,-90^\circ$

    \item In the lowest panel corresponding to $a_{\mu\tau}$, we observe that $2\sigma$ contours for $\delta_{13}=90^\circ,\phi_{\mu \tau}=90^\circ$ and $\delta_{13}=-90^\circ,\phi_{\mu \tau} =-90^\circ$ stretch over full range of $\phi_{\mu\tau}$. However, in the middle panel, very good precision is obtained for $\delta_{13}=0^\circ,\phi_{\mu \tau} =0^\circ$ with a closed $3\sigma$ contour.

    \item The best sensitivity for $\phi_{e\mu}$ is seen for $\phi_{e\mu}^{tr}=90^\circ,\delta_{13}^{tr}=90^\circ$, whereas best sensitivity for $\phi_{e\tau}$ is obtained at $\phi_{e\tau}^{tr}=0^\circ,\delta_{13}^{tr}=0^\circ$. 
\end{itemize}
In fig.~\ref{fig:chi-pre-t2hk}, we plot similar contours for T2HK. We argue that the precision in $\delta_{13}$, $\phi_{e\mu}$, $\phi_{e\tau}$ is poorer for T2HK. This is expected since at 295 km CP sensitivity is less. We can observe that the contours are similar for T2HK and T2HKK in $\delta_{13}-\phi_{\mu\tau}$ plane, i.e., precision is similar in both configurations also seen from fig.~\ref{fig:chi-amt}. The best sensitivity is obtained in T2HK also for  $\delta_{13}=0^\circ,\phi_{\mu\tau}=0^\circ$.

\section{\label{sec:result}Discussions}
The main focus of our work is to investigate the CP sensitivity in the future T2HK/T2HKK experiment in the presence of the CPT violating LIV parameters. We first study the CP discovery potential for individual baselines of 295km and 1100 km in the presence of LIV phases and ascertain the role played by neutrino and anti-neutrino contributions to the total $\chi^2$. Next, we obtain the sensitivity for T2HK and T2HKK configurations. We find that T2HKK gives a better sensitivity because of the synergistic effects of 295 km and 1100 km for LIV in the $e-\mu$ and $e-\tau$ sectors. We have identified synergy in parameters of $a_{\alpha\beta}, \theta_{23}, \phi_{\alpha\beta}, \delta_{13}$. However, for LIV in the $\mu-\tau$ sector, both configurations give similar sensitivity. This is because of the weak dependence of $P_{\mu e}$ on $\phi_{\mu\tau}$.

We also obtain the precision of $\delta_{CP},\phi_{\alpha,\beta}$ for various true values of these phases in T2HK, T2HKK. We have found that the sensitivity of $\delta_{13}$ is better for T2HKK configuration in the presence of $a_{e\mu},a_{e\tau}$. In the case of $a_{\mu\tau}$, the sensitivity is best for $\delta_{13}=0^\circ,\phi_{\mu\tau}=0^\circ$. 

\begin{acknowledgments}
SG acknowledges the J.C. Bose Fellowship (JCB/2020/000011) of the Science and Engineering Research Board of the Department of Science and Technology, Government of India. SG gratefully acknowledges support by the Fulbright Nehru Program, which is sponsored by the U.S. Department of State and United States India Education Foundation. Its contents are solely the responsibility of the author and do not necessarily represent the official views of the Fulbright Program, the Government of the United States , or the United States India Education Foundation. 
\end{acknowledgments}


\nocite{*}
\bibliographystyle{apsrev4-1}
\bibliography{ref_liv}

\end{document}